# A Simulation Model for Pedestrian Crowd Evacuation Based on Various AI Techniques

Danial A. Muhammed[1*], Soran A.M. Saeed[2], Tarik A. Rashid[3]

[1] Computer Department, College of Science, University of Sulaimani, Sulaymaniyah 46001, Kurdistan, Iraq
[2] Sulaimania Polytechnic University, Sulaymaniyah 46001, Kurdistan, Iraq
[3] Computer Science and Engineering, University of Kurdistan Hewler, Erbil 44001, Kurdistan, Iraq

Corresponding Author Email: danial.muhammed@univsul.edu.iq

*Keywords:*
*evacuation models, computational modeling, simulation, participants' emergency behavior, evacuation time, environment, engineering applications*

**ABSTRACT**

This paper attempts to design an intelligent simulation model for pedestrian crowd evacuation. For this purpose, the cellular automata (CA) was fully integrated with fuzzy logic, the kth nearest neighbors (KNN), and some statistical equations. In this model, each pedestrian was assigned a specific speed, according to his/her physical, biological and emotional features. The emergency behavior and evacuation efficiency of each pedestrian were evaluated by coupling his/her speed with various elements, such as environment, pedestrian distribution and familiarity with the exits. These elements all have great impacts on the evacuation process. Several experiments were carried out to verify the performance of the model in different emergency scenarios. The results show that the proposed model can predict the evacuation time and emergency behavior in various types of building interiors and pedestrian distributions. The research provides a good reference to the design of building evacuation systems.

## 1. INTRODUCTION

Occurring emergencies in big places makes the evacuation operation difficult due to adding confusion, fear and even vagueness and anxiety to mass residents [1]. Constructed environment, evacuees' communication, and the plans and measures in place to direct the reactions of participants to environmental conditions and characteristics can influence the evacuation system. Therefore, various evacuation methods appear, such as protective, preventive, rescue, and reconstructive evacuations [2]. The problem of evacuation is not solely the physical movement of evacuees; it is multifaceted and related to the physical and social circumstances, such as the high possibility for hazard, great stage of pressure, and inadequate data. These circumstances illustrate robust communication between environment, danger, egress process, population demographics, and participant behavior [3]. A crowd is a communication of a set of people [4]. Various features belong to crowds; therefore, during an emergency, a number of behaviors would be exposed in community areas [5], and simulation is a means of expecting behaviors via responding to the "what-if" situations [6]. Hence, Crowd evacuation simulation is a method for mimicking the flow of participants for the same goal of evacuation from the same place [4]. In the last two decades, rising dedicated attempts via researchers clearly have been seen to explore crowd evacuation during an emergency. Evacuation dynamics have been widely considered to eliminate deaths and injuries during emergency pedestrian incidences [7-13]. A vast number of models is developed to know individual behaviors, such as how individual react a specific condition and how to deal with varied kind of scenarios in the crowd during an emergency; the entire investigations have been there for creating a foundation of enhancement in managing emergencies [14-20]. Modeling is a way of answering difficulties, thus, a newly developed framework after analyzing and controlling via basic pro-test shows the authenticity of its procedures to be mentioned as a model [4]. Existing models can be separated into three main groups: the first one is classical models, the second one is hybridized models, and the third one is a generic model. Each group comprises some models, in the recent past; various applications to investigate crowd evacuation in both normal and emergency situations were done due to these models.

Microscopic is one of the classical models and the details of the individuals and individual behaviors are precisely identified. However, it has a problem in dealing with the large size of participants. In microscopic, cellular automata, lattice gas, social force, agent-based, game theory, and experimental approaches were designed. In the last two decades, cellular automata models have been created to consider an evacuating group of individuals under different circumstances. These models can be categorized into two groups. The first depends on the associations among situations and pedestrians. For example, in 2007, Varas et al. utilized a two-dimensional cellular automaton model to imitate the process of leaving from a single and double door classroom with complete ability. In this study, to each grid, the structure of the room, obstructions spreading, floor field were measured. Moreover, the effect of panic as a counted dimension was calculated which % 5 possibilities of not moving. The model applied random selection to cope with the collision issue. Therefore, the proposed model changed into non-deterministic via these characteristics. From the simulation result, it was clearly observed that the best locations of the door and evacuation efficiency were not enhanced by substituting a double door with two distinct doors. Finally, for the evacuation time, a number of persons and way out width were considered due to suggesting simple scaling law [21].

In 2011, Alizadeh proposed a CA model to analyze process of evacuation in a room contained obstructions and had different configuration of the room, such as exit and obstruction locations, width of the exit, light of the area, evacuee's psychology, and spreading of the crowd, which had a significant impact on the evacuation process. The case study of this model was the restaurant and a classroom. The impact of the distribution of the evacuees, location, and width of the door on time of the evacuation argued and output of the model was com-pared with some static models [22]. In 2014, Guo, Ren-Yong created a model based on CA with better separation of the area and advanced speed of walking to demonstrate moving away of pedestrians from a room with a single exit door. When experiments were simulated, near the exit they noticed the shape of the crowd was affected by both the separation of the area and advanced speed of walking, duration of the individuals at various positions, and efficiency of the evacuees expressed via two-time indicators. Besides, the association between width and flow of the exit was expressed via this model [23].

In 2015, Li and Han recommended a model for simulating pedestrian evacuation based on extended cellular automata to configure different behavioral propensities in people. Familiarity and aggressive were the two chosen behavioral propensities to be examined via this model. During the experimentations of this simulation, behavioral parameters and pedestrian flow arrangements were verified. The output of this study demonstrated that evacuation time de-creases with increasing individuals' familiarity and increases with the increasing individuals' aggressiveness. Thus, it is clearly seen that the best evacuation was based on the familiarity of the individual with avoiding aggressive behavior [24].

In 2018, Kontou et al. created a model of crowd evacuation based on cellular automata (CA) parallel computing tool to simulate and evaluate behaviors and distinct characteristics of the people within the area of evacuation, which comprises disable people. A secondary school in the region of Xanthi, which contained disable children, was chosen for the simulation process. With noticing and existing earthquake, the school controlled safety exercise; the overall time of the evacuation was recorded. Finally, the proposed model via the empirical data authenticated and there was a convenience inference to the special area [25]. Table 1 shows information for future use.

The main goals of this paper are: 1) Collecting most of the research works related to various applications applied to pedestrian evacuation using a cellular automata approach in the microscopic classical model. 2) Investigating features, techniques, and implications of the different applications. 3) Learning from 1 and 2 to make a decision to design a new intelligent and reliable model to simulate participants' appearing emergency behaviors and evacuation efficiency during an emergency evacuation. The model is designed based on an amalgamation of cellular automata, fuzzy logic technique, KNN algorithm, and some statistical equations. Consequently, this paper provides details and specifics in the area of pedestrian evacuation and provides opportunities to the researchers to reach the related information easily and determine their future research directions.

This research study's contributions are: 1) Utilizing participants' physical, biological, and emotional elements to apply different speeds for each participant. 2) Incorporating the speed with different elements, such as environment, participants' distributions, and familiarity of the participant to the exit doors to the evacuation model to improve their appearing emergency behaviors and evacuation efficiency.

The remainder of this paper is organized as follows: Section 2 presents the research method. Section 3 shows the result and analysis of the simulations. Finally, Section 4 concludes the main points and suggests future research studies.

## 2. RESEARCH METHOD

The following structure, which is presented in Figure 1 define subsections of the suggested framework of pedestrian evacuation through the first floor during the existing emergency to record each of the pedestrian evacuation time and emergency behaviors.

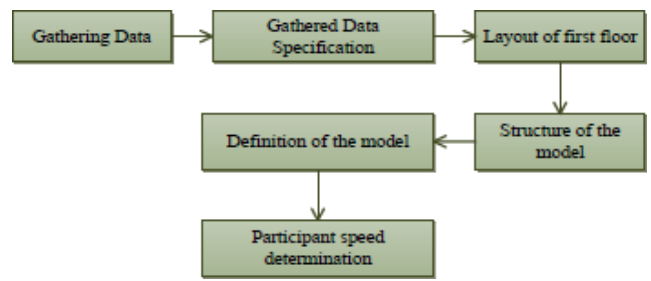

**Figure 1.** Proposed model methodology framework

### 2.1 Gathering data

The methodology of our framework started with gathering data. Essentially, different procedures are involved in gathering data. It is obvious direct observing of the situations appears within the evacuation process is vital to acquire reliable data for different forms of environments. However, due to difficulty in performing this method, new methods were applied, such as applying a general database to hold whole parameters' values with lacking information about provided agents [26]. In the crowd evacuation, the interior of the building plays a great role, besides; there isn't sufficient investigation for influences of the geometry inside the researches [27]. During the evacuation, obstructions' locations and size of the area have a great role [28]. Gathering data for our model is done via two common approaches; interviewing and questionnaire. In the interviewing, the participants faced some physical, biological, and emotional properties structured questions. Also, they asked a number of dichotomous questions of the type of closed-ended questions.

### 2.2 Gathered data specification

The collected data consists of 81 participants 35 males and 46 females. Furthermore, age and weight for the male participants were 18 to 50 years old and 68 to 95 k/g. Age and weight for the female participants were 18 to 43 years old and 57 to 83 k/g. This part of the gathering data utilized in defining the agent's speed while the second part of the collecting data was via the questionnaire was utilized in implementing the agent's behaviors, such as failing, waiting, helping, jumping, and others. Results of the 81 participants for the first part and second part with their analyzing results were briefed in Tables 2 and 3.

The percentage value in tables 2 and 3 represents the number of people with the behaviour over the number of participants, for example, in Table 2, the wait behaviour is 20, which means 20 people are waiting and the number of participants is 35, thus, 20/35 = %57.14.

Table 1. Highlighting the CA method and its features, techniques, and implications of current simulation models

| Type of models | References | Methods | models | occupants | State of simulation | Investigated occurrences and behaviors |
|---|---|---|---|---|---|---|
| | | | Cellular Automata Model Applications | | | |
| Classical Models | [21] | CAM | Microscopic | Homogenous | Emergency | Effect of obstacles |
| | [22] | CAM | | | Emergency | Impact of distribution of the evacuees, location, and width of the door on time of the evacuation argued |
| | [23] | CAM | | | Normal | The shape of the crowd, duration of the individuals at various positions, the efficiency of the evacuees expressed via two-time indicators, the association between width and flow of the exit |
| | [24] | CAM | | Heterogeneous | Normal | Familiarity and aggressive, evacuation time |
| | [25] | CAM | | | Emergency | Disable children, evacuation time |
| | | | proposed New Model | | | |
| proposed model | proposed model | CA (combined with the idea of fuzzy logic technique, KNN algorithm, and some statistical equations) | Microscopic | heterogeneous | Emergency | Determining heterogeneous speed to evacuees based on their properties, effect of environment, evacuees distribution, number of room exit doors and main exit doors, and familiarity of evacuees to the exit door on evacuation time and appeared emergency behaviors within the evacuation process |

Table 2. Questionnaire of 81 participants for their behaviors during an emergency evacuation

| Gender | Number of participants | Behaviors during an emergency evacuation | | | | |
|---|---|---|---|---|---|---|
| | | Wait | Aside | Jump over | Help | Wai to fail the person |
| Male | 35 | 20 (%57.14) | 24 (%68.57) | 13 (%37.14) | 10 (%28.57) | 18 (%51.43) |
| Female | 46 | 18 (%39.13) | 40 (%86.96) | 5 (%10.86) | 3 (%6.52) | 6 (%13.04) |

Table 3. Interview of 81 participants' for their physical biological and emotional properties

| Gender | Number of participants | Physical and Biological Properties | | | | Emotional Properties | |
|---|---|---|---|---|---|---|---|
| | | Age(yrs.) | Weight(kg) | Levels | Disease | Shock | Collaboration |
| Male | 35 | 18 – 50 | 65 – 95 | V. Low | 25 (%71.43) | 18 (%51.43) | 26 (%74.29) |
| | | | | Low | 5 (%14.29) | 8 (%22.86) | 5 (%14.29) |
| | | | | Medium | 2 (%5.71) | 4 (%11.43) | 2 (%5.71) |
| | | | | High | 2 (%5.71) | 3 (%8.57) | 1 (%2.86) |
| | | | | V. High | 1 (%2.86) | 2 (%5.71) | 1 (%2.86) |
| Female | 46 | 18 – 43 | 57 – 83 | V. Low | 35 (%76.09) | 5 (%10.87) | 21 (%45.65) |
| | | | | Low | 5 (%10.87) | 5 (%10.87) | 13 (%28.26) |
| | | | | Medium | 3 (%6.52) | 18 (%39.13) | 7 (%15.22) |
| | | | | High | 2 (%4.35) | 15 (%32.61) | 3 (%6.52) |
| | | | | V. High | 1 (%2.17) | 3 (%6.52) | 2 (%4.35) |

**2.3 Layout of the first floor**

First floor layout for our presented framework was the first floor of one of the institutes' in Sulaimaniyah city in the Kurdistan region of Iraq, which was a polytechnic institute. The first floor of this building contains a cafeteria, which different people visit this part of the building such as, employees, students, visitors, and cafeteria staff with various health, weight, age, and different response or behavior throughout the emergency case. Inside the first floor, 540m² is provided, for the cafeteria with a width of 36 meters and length 15 meters. This area is partitioned into three different subareas; students' area, employees' area, and kitchen and services areas. The area of employees is 10 meters width and 15 meters length, inside the employees' area there are 6 tables, and one exit doors with 2.5 meters width. The area of students is 17 meters width and 15 meters length, inside the students' area, there are 8 tables and two exit doors with 2.5 meters width for each. Area of kitchen and services is 9 meters width and 15 meters length; this part contains some components for cooking, refrigerator, and cooler and it has two exit doors with 2 meters width, one is on the students' area and the other one from the back end of the area. Figure 2 illustrates the outline of a part of the first-floor polytechnic institute building.

**2.4 Structure of the model**

A structure is designed for this framework and within the design, different characteristics, such as environment, individuals, group behaviors, and emergencies are mentioned. For the first characteristic, the size of the area, the boundary of the floor, the size of the exit doors with their locations, and obstacles are determined. For the second characteristic, agent position, group behaviors such as fail, wait, help, and jump is determined and individual proper-ties such as age, weight, gender, and others are specified to define various speeds of agents. Furthermore, the third characteristic is emergency situations such as fire, smoke, and others are mentioned and participated in defining agents speed with the individual properties in the second characteristic. After these determinations, environment components, agent speed, and behavior patterns together are employed to execute the evacuation process and then evacuation time and various appeared behaviors of each agent during the evacuation process would be recorded. Figure 3 shows the structure of the presented model.

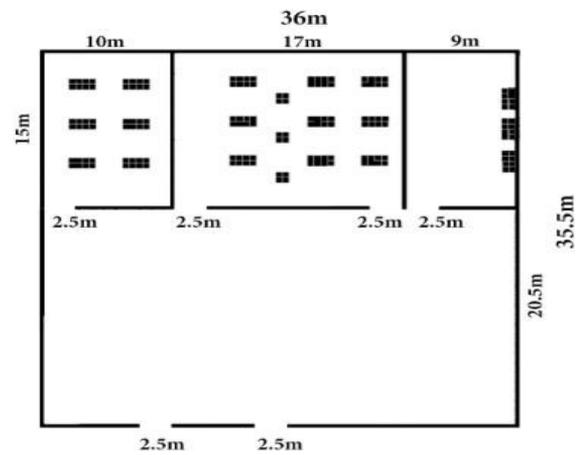

**Figure 2.** Illustrate the outline of a part of first-floor polytechnic university's precedency building

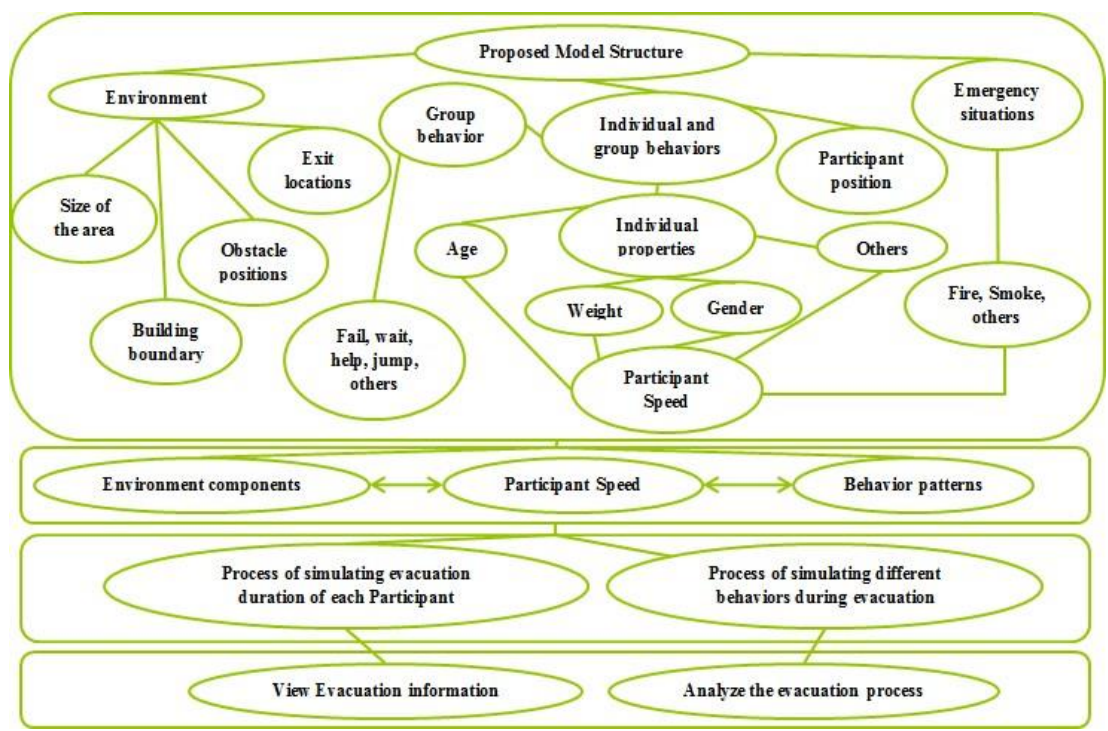

**Figure 3.** Illustrates the proposed model structure

## 2.5 Definition of the model

In the definition of this framework, we demonstrate the individuals, individuals' behaviors, case study's geometry, and spreading of individuals throughout the geometry. This framework for the pedestrian evacuation during emergency built based on cellular automata (CA) approach, which is one of the approaches of the microscopic classical model, and combining it with the idea of fuzzy logic technique, some statistical equations, and K-Nearest Neighbor (KNN) algorithm. This floor is divided into grids, while the CA method has many characteristics such as discrete, continues, and dynamic [7]. Size of each cell set with $0.5*0.5m^2$. Naturally, each cell might or might not be available due to the individual, exit doors, walls or obstacles. A two-dimensional array with the dimensions of the floor is used to separate the area into grids. Inside this framework, males and females with various properties, for instance, emotional, biological, and physical properties are applied. Spreading of individuals within this floor of the case study inside this framework is either randomly or manually. Expressing an emergency in this study is not visible for the agents through the evacuation, but it is due to a coefficient, which puts an impact on the speed of the individuals. Different components implemented to construct this framework, such as individuals with yellow color, obstacles with gray color, and exit doors with dark blue color, rooms' wall with pink color, and main walls with the cyan color, which determine the area of the evacuation. Figure 4 (a) illuminates various implemented components inside the proposed framework. Individuals can move with 8 directions to evacuate the building. Figure 4 (b) is an individuals' movement with 8 directions.

During the evacuation process, individuals evacuate with determined speed and due to the strike with other individual dilation occurs for some while. In this framework, some behavior is implemented, for instance, when an individual reaches an obstacle, distance from both ends of the obstacle would be checked by the individual, and then decides to pass the obstacle from the nearest one. Figure 4 (c) is the participant's behavior when faced with an obstacle.

Colliding individuals in this framework causes to collider to make a decision to go to another empty location and pass the other collider or wait to the other collider to move from the occupied location. For the passing situation, the framework changes the individual's color from yellow into light green, while for the waiting situation the framework changes color from yellow into dark green. Figure 4 (d) is the participant's behavior when faced with other participants.

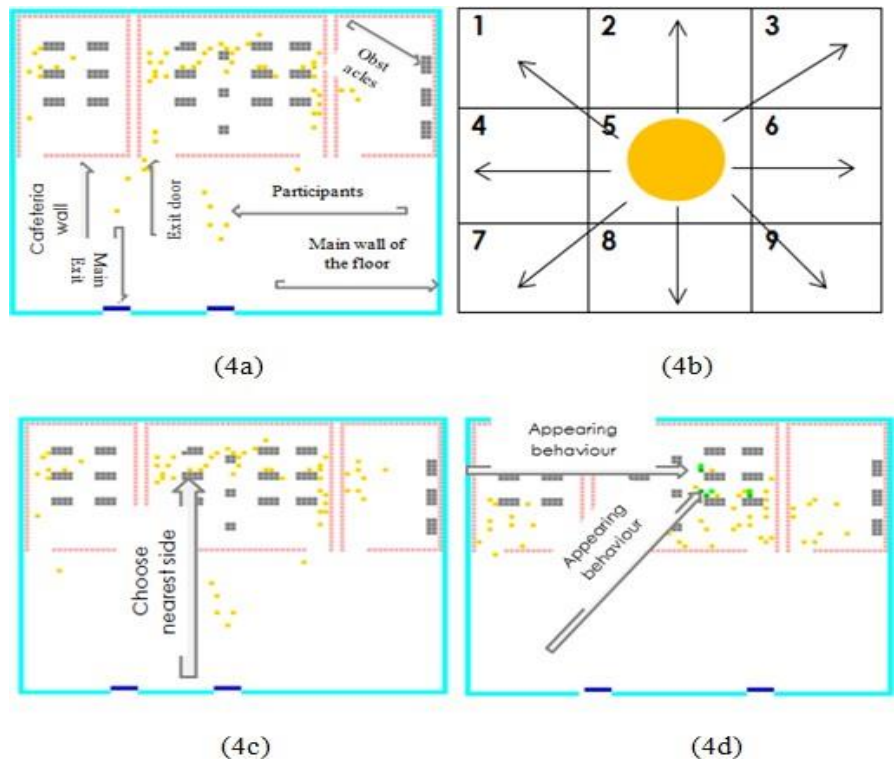

**Figure 4.** Model definition: (a) various implemented components inside the proposed framework, (b) Individuals' movement with 8 directions, (c) participant's behavior when a faced obstacle, (d) participant's behavior when faced other participants

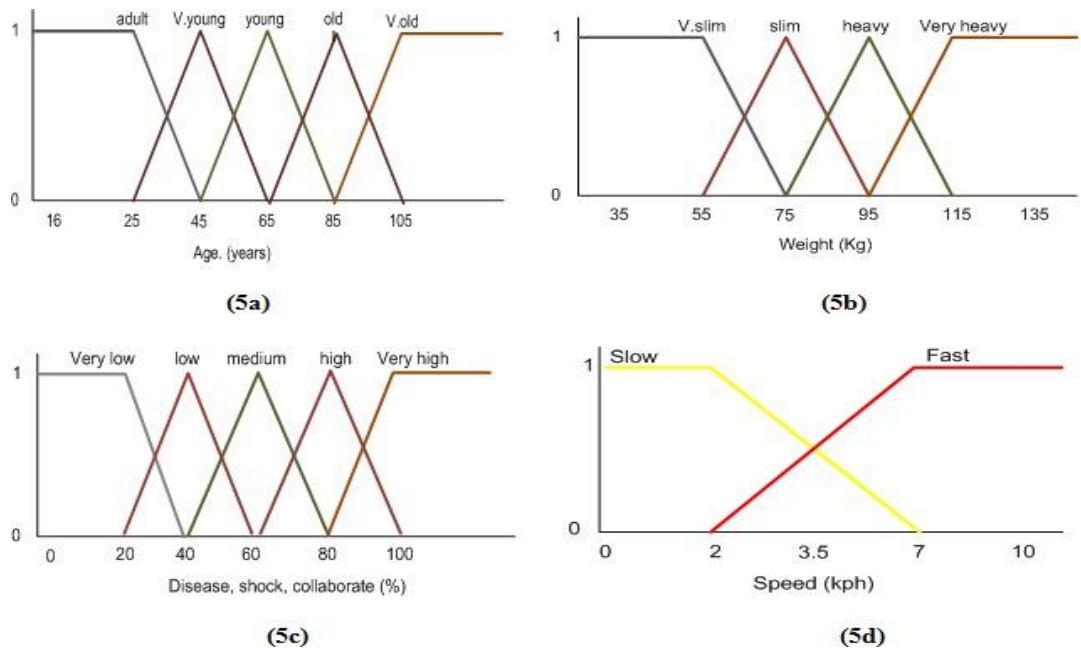

**Figure 5.** Model definition: (a) Age membership function, (b) Weight membership function, (c) Disease, shock, collaborate membership function, (d) Speed membership function

### 2.6 Participant speed determination

Inside this framework, determination of individual speed with considering the mentioned individual properties such as physical, biological, and emotional was due to taking benefits from the idea of fuzzy logic technique. By means of accepting

287

the fuzzy logic idea to the fuzziness of each parameter, this framework reaches a realistic solution. Many-valued logic is the output of reality, which managed in fuzzy logic [29]. This framework is designed based on the fuzzy logic idea, thus it examines the individual properties that found from the gathered data and creates fuzzy linguistic variables represent the qualities spanning a particular range. For example, disease {very low, low, medium, high, very high}, weight {very slim, slim, heavy, very heavy}, age {adult, very young, young, old, very old}, collaboration {very low, low, medium, high, very high} and shock {very low, low, medium, high, very high}. Then the framework creates a membership in each assortment. Figures 5 (a, b, c, d) present the membership functions of the age, weight, disease, shock, collaborate separately.

The following example clarifies how evacuees speed is determined in our model:

After collecting and analyzing individual properties, the designed membership functions (See Figures 5 (b), 5 (c), 5 (d) and 7) are applied to identify the degree of the properties of an individual with its speed range. The degree comprises of two values; lower value and upper value, and then weighted mean equation (See Eq. (1)) is used to work on these values [30].

$$weightprop = \frac{\sum_{i=1}^{n} pw_i * srd_i}{\sum_{i=1}^{n} pw_i} \quad (1)$$

$pw_i$ is the degree of the given property's weight and $srd_i$ is the speed range for the degree. In this study, equation (1) has different forms (see Eq. (2) and Eq. (3)) to make heterogeneity inside a single class interval. An individual with age and weight properties, their properties' weights defined separately with applying Eq. (2) and Eq. (3).

$$weightprop = weighted\ mean = (lo * minisrd + up * maxisrd)/(lo + up) \quad (2)$$

$$weightprop = weighted\ mean = (up * minisrd + lo * maxisrd)/(lo + up) \quad (3)$$

From Eq. (2) and Eq. (3), $lo$ is lower value and $up$ is upper value, $minisrd$ is minimum interval speed range, and $maxisrd$ is maximum interval speed range of the mentioned properties. Before using these two equations, designed memberships (See Figure 5 (a) and 5 (b)) need to be used to find the degree of the given properties. For example, if the individual is 38 years old, designed membership for age property is used to find the degree of that given age property (lower value and upper value of the age). To find the right class interval, the model checks the intervals and finds the right one (See Figure 7). Hence, the framework determines the desired value for 38 years old is 0.35 adult and 0.65 very young. On the other hand, to find the property at which speed range is, the designed membership function of speed is used (See Figure 5 (d)). The specified range for speed in this framework is between 2k/h and 7k/h. This framework breaks down the speed range into several class intervals. This separation is for making logical heterogeneity in values of property's weight for various individuals within the speed range. For example, for an individual with 38 years old class interval of speed is 5k/h - 6k/h, whereas for an individual with 46 years old is 4k/h - 5k/h (See Figure 7). The aim of this diversity in the class interval for speed is associated with logic; younger persons are faster than elder persons. Additionally, finding the middle of the chosen class interval is necessary for finding mid value, Midvalue equation (See Eq. (4)) [30] would be utilized.

$$Midvalue = lov + upv/2 \quad (4)$$

From Eq. (4), for the chosen class, the lower limit is denoted with $lov$ and the upper limit is denoted with $upv$. The class interval would be divided into two distinct parts via Midvalue; first half of the interval and second half of the interval (See Figure 7). This division makes the values of given properties' weight different in the two distinct halves of the interval, which leads to remove weight redundancy and arrange different properties weights in both halves of the interval. The result of the Midvalue equation and the given property's value make the framework decide to use Eq. (2) or Eq. (3). For example, when the given property value is greater than the Midvalue result, the framework applies Eq. (3), conversely, when the given property value smaller than the Midvalue result, the framework applies Eq. (2). Both Eq. (2) and Eq. (3) could be used when given value for the property is equal to the Midvalue result. The same operations are done to find the weight of the property of weight. Results of Eq. (2) and Eq. (3) are used to calculate properties' weights as in Eq. (5), which is the mean equation [30] after combination the equation with gender factor $gen_i$ and emergency coefficient $em_i$.

$$desiredSpeed = \frac{(\sum_{i=1}^{n} pw_i)}{n} * g_{eni} * e_{mi} \quad (5)$$

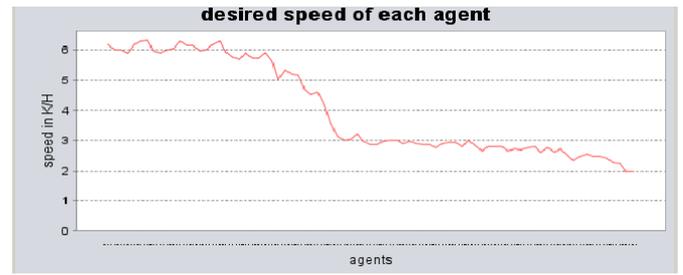

**Figure 6.** Result of determining the desired speed for 81 participants based their properties

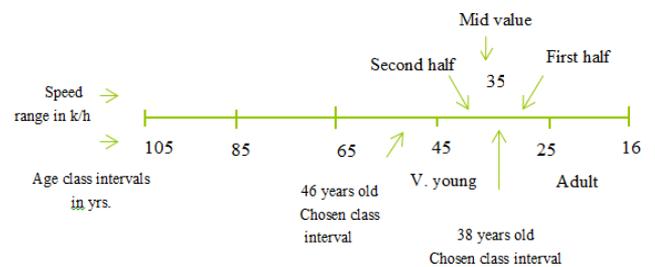

**Figure 7.** How weights of participant's properties are identified

Finally, gender and emergency factors were counted for the desired speed. From this, the desired speed is affected by those factors, while speeds for male and female is different even both have the same properties. The female speed in our expectation for the simulation was decreased by 0.5. Thus, if the speed of males is 6 kph, it is for female agents decreased to 3 kph. On the other hand, during an emergency, evacuees run as fast as they could. Hence, the speed of the evacuees during an emergency is more than the normal situation. Figure 6 illustrates the conducted desired Speed in kph for the agents.

## 3. RESULTS AND ANALYSIS

In this research work, the model is based on various scenarios, which have been written and executed through various experiments. Concise of experiments' results for the scenarios via the developed intelligent model were briefed inside Table 4. For this simulation process, data gathered for 81individuals. As mentioned in the previous sections within the gathering data, individual properties were recorded for each of them. Therefore, these properties were entered into the model, based on these properties individuals were generated and desired speeds were specified for them. These individuals arbitrarily set through the cafeteria. Figure 8 illustrates the participants' distribution through the first floor.

The above-mentioned scenarios could be briefed in the following points: as mentioned in subsection 4.2, 43.21 percent was male and 56.79 percent was female with (1) One main exit for the first floor, two exit doors for students part, only one exit door to each employee part and staff part, and distribution of the individuals in small and large area of the cafeteria. (2) Two main exit doors for the first floor, only one exit door for each of the student part, employee part and staff part, and distribution of the individuals in a small and large area of the cafeteria. (3) Two main exit doors for the first floor, two exit doors for student part, one exit door for employee part and staff part, and distribution of the individuals in a small or large area of the cafeteria. (4) Three main exit doors, two exit or one exit doors for students' part, one exit door for each employee part and staff part, and distribution of the individuals through a large area of the cafeteria. Individuals' familiarity is considered with the points (2), (3) and (4).

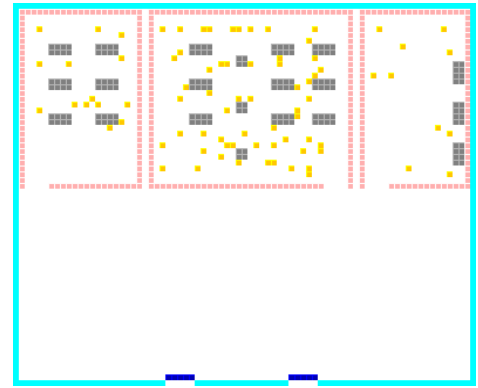

**Figure 8.** Illustrates the participants' distribution through the cafeteria

**Table 4.** Experimentations' results from the developed model

| Experimentations | Scenarios | Rate of C.W.A (%) | Rate of C.W.O (%) | O.B.D.E | Evacuation time min:sec:ms |
|---|---|---|---|---|---|
| No1A | Evacuees distributed in an average small area in the cafeteria, two main exits, each of the employee part, student part, and staff part has one exit door, evacuees were no familiar with the exits | 61.728 | 30.864 | Aside, wait | 1:0:941 |
| No1B | Same as the No1A but evacuees were familiar with the exits | 51.852 | 65.432 | Aside, wait | 0:47:261 |
| No1C | Same as the No1A but evacuees distributed in a larger area of the cafeteria | 27.16 | 20.988 | Aside, wait, help | 0:46:711 |
| No1D | Evacuees distributed in an average small area in the cafeteria, two main exits, two exit doors for students part, and each of the employee part, and staff part has one exit door, evacuees were familiar with the exits. | 17.284 | 33.333 | Aside, wait | 0:47:130 |
| No2A | Only one main exit door was used, evacuees distributed in a larger area of the cafeteria | 32.099 | 13.58 | Aside, wait | 0:41:635 |
| No2B | Same as No2A, but all agents distributed through a very small portion of the floor | 54.321 | 20.988 | Aside, wait | 0:50:682 |
| No2C | Same as No2A but agents positions were near the exit door | 49.383 | 4.938 | Aside, wait, help, jump over | 0:35:373 |
| No2D | Only one main exit door was used with two exit doors of student part and employees' part and staff part has only one exit door, evacuees distributed in a small area of the cafeteria | 44.444 | 17.284 | Aside, wait | 0:43:487 |
| No3A | More than two main exit doors and increase evacuees' distribution area | 23.457 | 13.58 | Aside, wait | 0:41:241 |
| No3B | Same condition No3A, they are all familiar with the exit doors | 16.49 | 16.049 | Aside, wait | 0:37:976 |
| No4A | Two exit doors for the student part, three main exit doors and evacuees were no familiar with the exits | 25.926 | 7.407 | Aside, wait | 0:40:51 |
| No4B | Same condition No4A, evacuees familiar with the exits | 22.222 | 7.407 | Aside, wait | 0:31:307 |

The simulation results of the above-mentioned points via our developed intelligent model were briefed in Table 4. Some abbreviations were used, such as C.W.A is a collision with agents, C.W.O is a collision with obstacles, and O.B.D.E is occurring different behaviors during evacuation. Inside the developed our new model, it was appeared that increasing number of room exit doors, main exit doors with familiarity of the evacuees to the exit doors and distribution in a larger area of the floor are the key factors to minimize the congestion, collision and appearing agent's emergency behaviors that directly have a great influence on improving evacuation efficiency. In the remaining of this section, the result of our model simulation is discussed. For details of simulation results, refer to Supplementary File comprises 36 pages is available in the supplementary Files.

Obviously specifying heterogeneous speed for the participants creates jamming due to colliding agents with each other and with obstructions. This situation was clearly seen in the simulation results. Furthermore, changing numbers of main exit doors and cafeteria's distinct parts exit doors create variation in evacuation time and appearing emergency behaviors, such as C.W.A, C.W.O, and O.B.D.E. For example, in experimentation of No1C, No2A, they had different numbers of main exit doors, but they had the same distributions, C.W.A, C.W.O, and O.B.D.E were recorded as 27.66 percent, 33.33 percent, and (aside, wait) for No1C respectively, 45.679 percent, 50.617 per-cent, and (aside, wait, help) for No2A respectively. Besides, it was found that with adding a number of main exit doors C.W.A and C.W.O were decreased and different O.B.D.E appeared. Furthermore, agents' evacuation times were reduced and recorded 0:46:74 for No1C, 0:47:805 for No2A, the time managed as a format of min.sec.ms.

When participants distributed through a larger area of the cafeteria great effect would be created on the occurring behaviors and also on the evacuation time. For instance, in the experimentation of No1A and No1C behaviors, such as C.W.A, C.W.O, OBOE were 61.78, 30.64, and (aside, wait) for No1A respectively, while these behaviors were 27.859, 20.988, and (aside, wait, help) for No1C, respectively.

Because of the larger distribution in No1C collision behaviors significantly decreased between participants and participants with obstacles. Besides that, evacuation time considerably improved and recorded as 1:0:941 for No1A and 0:46:711 for No1C. On the other hand, participant familiarity with the exits doors had a great impact on changing the participants' evacuation behaviors and evacuation time, for instance, in the experimentations of No1A vs No1B, No3A vs No3B, and No4A vs no4B behaviors and evacuation time, such as, C.W.A and C.W.O changed from 61.728 to 51.852, 65.432 to 30.864, 1:0:941 to 0:47:261 for No1A and No1B, respectively, 23.457 to 16.49, 13.58 to 16.49, 0:41:241 to 0:37:976 for No3A and No3B receptively , and 25.926 to 22.222, 7.407 to 7.407 ,0:40:51 to 0:31:307, for No4A &No4B, respectively.

Furthermore, changing the number of exit doors made a great change in participant's evacuation behaviors and evacuation time, for example, in the experimentation of No3A vs No1A could be clearly noticed. In No1A which had two main exits and participants distributed through a small portion of the cafeteria, the appeared behaviors, such as, C.W.A, C.W.O, and O.B.D.E significantly more occurred than in No3A which had three main exits and participants distributed through larger area of the cafeteria, the result of this experimentation for behaviors, such as, C.W.A, C.W.O, OBDE, and evacuation time was recorded as 61.728, 30.864, (aside, wait), and 1:0:941 for No1A, receptively, 23.457, 13.58, (aside, wait), and 0:41:241 for No3A, respectively. The result shows appearing behaviors decreased and also evacuation time considerably improved due to an increase in the number of main exits and area of distribution.

From the experimentation result of No1D vs No2D, it appeared that increasing the number of exits is not the only reason to decrease appearing emergency behaviors and improving evacuation efficiency. For example, in the experimentation of No1D, there were two main exit doors, and in the experimentation of No2D, there was only one main exit door. However, the C.W.A is better to be compared to C.W.A in No2D, but C.W.O and evacuation time was better in No2D due to larger distribution.

## 4. CONCLUSION

Despite the fact that developed models based on pedestrian evacuation approaches have been extensively used and confirmed, there is still essential to create models that carefully simulate people's evacuation time and appeared behaviors during an emergency evacuation. It is vital to review many previous works and notice different factors that made an impact on participant behaviors during the evacuation process and also on participants' evacuation time.

In this study, an intelligent model built based on analyzing the previous applications of the cellular automata approach which is one of the microscopic model's approaches. Inside this model, cellular automata (CA) employed with a combination of the fuzzy logic idea to define the value of the participant's properties, such as physical, biological, and emotional and then defined values of the participant's properties were used within some statistical equation to define desired speed for that participant. Additionally, the KNN algorithm was applied to find the nearest exit door during the evacuation process.

This study discusses the impact of the defined desired heterogeneous speed for participants with a combination of participants' behavior, familiarity, environment, and participants' distribution through the distinct parts of the cafeteria on the appearing emergency behaviors and evocation efficiency. From simulation results appeared different properties of participants caused different speeds for participants. Hence, a collision between participants C.W.A occurred during the evacuation process. Different behaviors, such as C.W.O and O.B.D.E appeared and different evacuation time was recorded for each participant. More experiments presented that environment, such as obstacles, size and number of main and distinct parts of the cafeteria exit doors had a great impact on changing participants' emergency behaviors and evacuation time.

This research confirms that participants with different properties, various distributions through the cafeteria, and the presence of obstacles are the key to appearing in various behaviors and congestion. Conversely, it is approved that various designs may decrease the impact of those factors. Firstly, instead of one main exit use two or more main exit doors. Secondly, the distinct part of the cafeteria increase the number of exit doors, meanwhile, changes the distribution of the participants inside the cafeteria into larger distribution. Thirdly, introduce the exit doors to the participants to choose the closest exit to evacuate as these situations simulated with the proposed model.

In the future, designing and implementing fire will be examined. Other features, such as: how the fire spread out through the building according to its environment will be added. Additionally, the effect of the fire on the agents' behaviors would be analyzed and discussed. References [31] is very close to our future work while in our work numbers of dead, injured or suffocated agents as a result of fire and smoke will be recorded. Additionally, this model will be enhanced to make a simulation for the second floor and above. Also, it can be seen as a real-world application and will be optimized by different optimizer algorithms, such as fitness dependent optimizer (FDO) [32], WOA-BAT optimization [33], donkey and smuggler optimization [34], and modified grey wolf optimizer [35] to find best location of the main exit door through the area of evacuation in a building. Finally, interested readers can read references [36-38] for the possibility of obtaining future directions.

## ACKNOWLEDGEMENTS

The study is fully funded by the University of Sulaimani (UOS). The authors would like to thank the UOS for providing facilities and equipment for this review work.